# Organic electrochemical neurons and synapses with ion mediated spiking

Padinhare Cholakkal Harikesh[1,6], Chi-Yuan Yang[1,6], Deyu Tu[1], Jennifer Y. Gerasimov[1], Abdul Manan Dar[1], Adam Armada-Moreira[1], Matteo Massetti[1], Renee Kroon[1], David Bliman[2], Roger Olsson[2,3], Eleni Stavrinidou[1,4], Magnus Berggren[1,4,5] & Simone Fabiano[1,4,5 ✉]

Future brain-machine interfaces, prosthetics, and intelligent soft robotics will require integrating artificial neuromorphic devices with biological systems. Due to their poor biocompatibility, circuit complexity, low energy efficiency, and operating principles fundamentally different from the ion signal modulation of biology, traditional Silicon-based neuromorphic implementations have limited bio-integration potential. Here, we report the first organic electrochemical neurons (OECNs) with ion-modulated spiking, based on all-printed complementary organic electrochemical transistors. We demonstrate facile bio-integration of OECNs with Venus Flytrap (*Dionaea muscipula*) to induce lobe closure upon input stimuli. The OECNs can also be integrated with all-printed organic electrochemical synapses (OECSs), exhibiting short-term plasticity with paired-pulse facilitation and long-term plasticity with retention >1000 s, facilitating Hebbian learning. These soft and flexible OECNs operate below 0.6 V and respond to multiple stimuli, defining a new vista for localized artificial neuronal systems possible to integrate with bio-signaling systems of plants, invertebrates, and vertebrates.

[1] Laboratory of Organic Electronics, Department of Science and Technology, Linköping University, SE-601 74 Norrköping, Sweden. [2] Department of Chemistry and Molecular Biology, University of Gothenburg, SE-412 96 Gothenburg, Sweden. [3] Chemical Biology and Therapeutics, Department of Experimental Medical Science, Lund University, SE-221 84 Lund, Sweden. [4] Wallenberg Wood Science Center, Linköping University, SE-601 74 Norrköping, Sweden. [5] n-Ink AB, Teknikringen 7, SE-583 30 Linköping, Sweden. [6] These authors contributed equally: Padinhare Cholakkal Harikesh, Chi-Yuan Yang. ✉email: simone.fabiano@liu.se

Research advancement in brain-machine interfaces, implantable/wearable devices, prosthetics, and intelligent soft robotics calls for close interaction between and integration of technology into nature. Since the fundamental building elements of life differ significantly from those utilized in electronic devices, the ability to link an artificial device with a biological system is crucial to the success of these domains. Neuromorphic systems[1] that borrow design concepts from biological signaling systems promise to bridge this gap. Although several software-based neuromorphic algorithms have been integrated into biomedical systems[2,3], hardware-based systems[4] that intimately interface with living tissues, evolve its function based on biological feedback[5,6], and utilize event-based sensing[7,8], and processing capabilities of the biological systems are ultimately necessary. However, circuits and devices made of silicon (Si)[7,9–11], commonly used in hardware neural networks and neural interfaces, suffer from several drawbacks such as rigidity, poor biocompatibility, the requirement for numerous circuit elements, and operation mechanisms that are fundamentally orthogonal to those of biological systems, making bio-integration difficult.

On the other hand, organic semiconductors are becoming a competitive alternative in this field, as seen by their growing applications in artificial synapses[12–17], nervetronics[18,19], and neural interfaces[20–22]. Their structural kinship with biomolecules makes them ideal for bioelectronic applications[23,24]. Organic semiconductors are solution-processable, biocompatible, biodegradable, soft, and conformable from a structural standpoint. Furthermore, they can also be easily functionalized to offer specialized excitation, sensing, and actuation capabilities, and they support the transport of both the electronic and ionic signals[25,26]. Since biological neurons communicate with each other through a regulated flux and polarization of ionic species, organic materials are an obvious choice for creating devices that replicate biological activities due to their ability to couple ionic polarization to the modulation of electronic charge transport.

Despite the success of organic materials in emulating neuromorphic functions and artificial nerves, there have been limited attempts to fabricate and bio-integrate artificial neurons, which are essential to enable spike-based information encoding that closely mirrors the processing strategies used by biological systems. Recently reported artificial neurons based on organic field-effect transistors (OFETs)[27] are promising in this regard. However, they require high voltage (5V) inputs for operation, which is an obvious critical issue when integrating with biology. In addition, as in the case of Si, their operation mechanism is fundamentally different from the ion-based mechanisms found in biological systems, making bio-integration, sensing, and response feedback difficult.

Organic electrochemical transistors (OECTs), which are modulated by a gate-driven ionic doping/de-doping of the organic bulk channel material[25], resemble the ion-driven processes and dynamics of biological systems. Compared to OFETs, OECTs operate at considerably lower voltages (<1 V), have a higher transconductance, maintain excellent threshold voltage stability, demonstrate strong ion concentration-dependent switching properties[28], and are in general highly biocompatible[29]. Aside from their use as chemical, physical, and biochemical sensors, they have been implemented as artificial synaptic devices[12,14,16,17,30] that exhibit both long and short-term plasticity. The robust device architecture and the solubility of the organic materials in benign solvents have also enabled the facile fabrication of large-scale printed OECT-based digital circuits[31–33]. These properties make OECTs the ideal candidates for developing printed, biocompatible artificial spiking neural circuits with ion-mediated spiking mechanisms closely resembling the signaling characteristics of biological systems.

Here, we report the first organic electrochemical neurons (OECNs), based on all-printed complementary OECTs. The OECNs exhibit several neuronal characteristics, including ionic concentration-dependent spiking, and spike-timing-dependent plasticity (STDP) on integration with printed organic electrochemical synapses (OECSs). It responds to a wide range of input currents (0.1–10 µA), resulting in frequency modulation of over 450%. For the first time, we utilize the ionic concentration-dependent switching characteristics of the transistor to modulate the frequency of spiking to a large extent analogous to biological systems and demonstrate its integration with Venus Flytrap (*Dionaea muscipula*). The electrochemical transistors enable chemical modulation of the spiking behavior, which is impossible in OFET-based or Si-based CMOS neurons. The all-printed OECSs, which form and evolve in operando by electropolymerizing the monomer precursor of a polymeric mixed ion-electron conductor, exhibits two different modes of operation based on ionic doping and electropolymerization-induced conductivity modulation. This dual-mode of conductance modulation enables short-term plasticity with paired-pulse facilitation and long-term plasticity with retention of over 1000 seconds, facilitating symmetric Hebbian learning upon integration with the OECN. We anticipate that the OECNs' soft nature, ability to be printed on flexible substrates, ion-modulated spiking and multi-stimuli response will open new avenues for facile integration with biological neural networks and applications in event-based sensors.

## Results

**Printed organic electrochemical transistors.** We chose the Axon-Hillock (A-H) circuit[1] (Fig. 1b) to fabricate leaky integrate and fire (LIF) type spiking OECN because it is the most compact model suitable for spiking neural networks (SNNs) and event-based sensors. The components of the circuit, made of n-type and p-type OECTs, are the Amplifying block and the resetting transistor $T_{reset}$ (Fig. 1b, c). The complementary OECTs (c-OECT) – made of the hole-transporting glycolated polythiophene (P(g$_4$2T-T), p-type) and the electron-transporting poly(benzimidazobenzophenanthroline) (BBL, n-type) – were fabricated by a combination of screen printing and spray coating techniques, as described in the Methods section (Fig. 1d). The OECTs have lateral Ag/AgCl gate configuration with screen-printed carbon and silver electrodes on a polyethylene terephthalate (PET) substrate. The silver underlayer decreases the line resistance while the carbon acts as a chemically inert layer to contact the polymer semiconductor. Owing to the volumetric bulk conductivity of the OECT channel material, it is easier to achieve a balanced driving strength between the p-type and n-type transistors by modulating the thickness of the semiconductor layer as opposed to complementary OFET-based circuits where the channel width/length of transistors must be altered to achieve the same. With the same device geometry ($W = 2000$ µm, $L = 200$ µm), the thicknesses of P(g$_4$2T-T) and BBL are 20 nm and 250 nm, respectively, to obtain the same driving strength (ON-state current ~50 µA at $V_{GS} = V_{DS} = \pm 0.6$ V) as shown in Supplementary Figure 1. A c-OECT inverter, comprising a pair of P(g$_4$2T-T) and BBL OECTs, is made to build the amplifying block in the A-H circuit (Supplementary Figure 2). The maximum DC voltage gain of the inverter is 26 (V/V) and we can reach nearly 200 (V/V) by cascading two inverters to form the amplifying block[34]. The fully printed OECN is shown in Fig. 1b and a typical spike firing of printed OECNs is presented in Fig. 1f.

**Organic electrochemical neurons, analogy with the biological neuron and bio-integration.** The operational mechanism of OECNs resembles that of the biological nerve cell[35,36] (Fig. 1a).

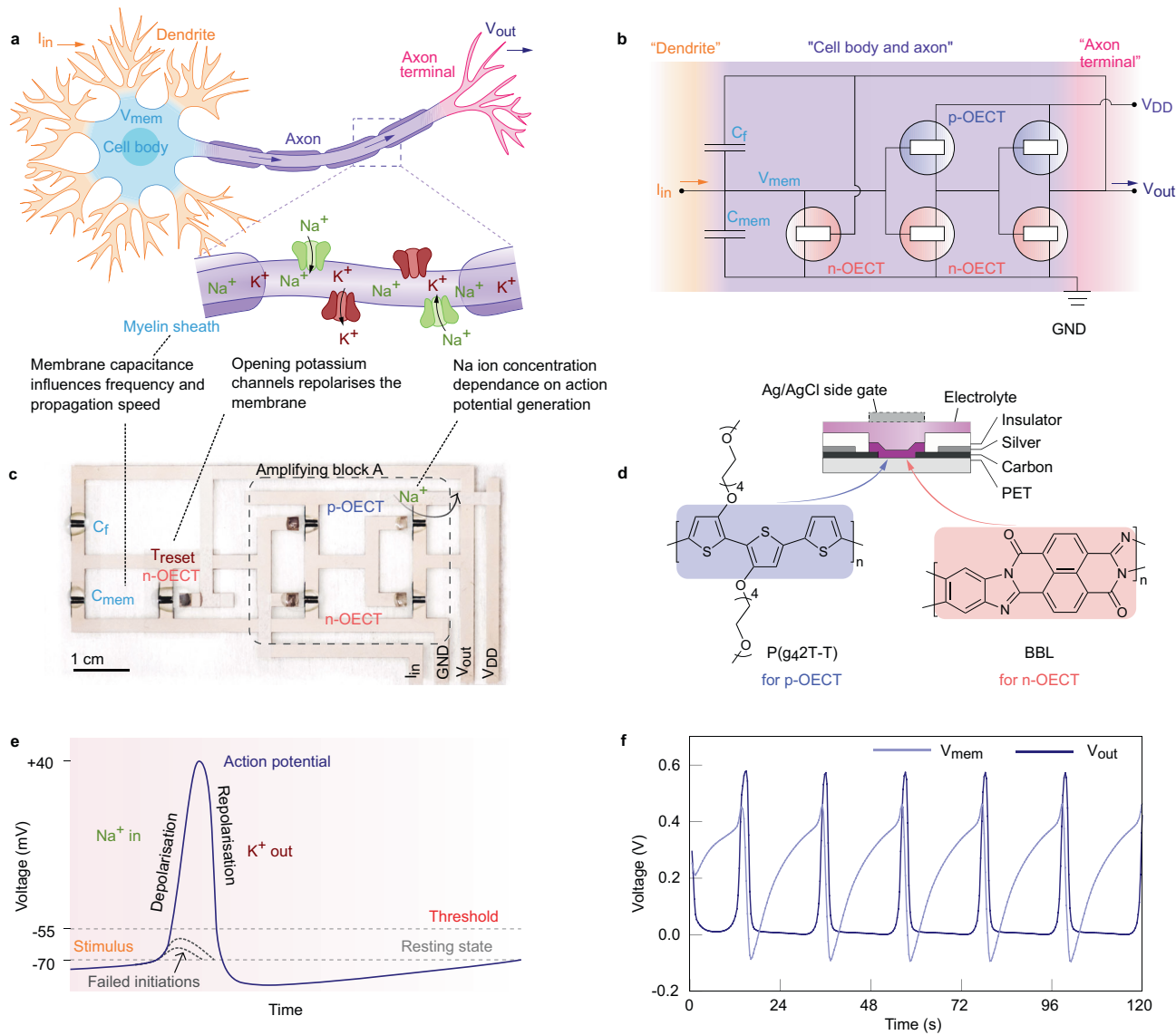

**Fig. 1 Organic electrochemical neurons and analogy with the biological neurons. a, b** Schematic of the biological neuron and its analogy with the organic electrochemical neuron based on Axon Hillock circuit. **c** Fully printed organic electrochemical neurons. **d** Structure of the printed p-type P(g42T-T) and n-type BBL OECTs. **e** Different phases of the action potential in a nerve cell. **f** Spiking behaviors of the organic electrochemical neurons with $C_{mem} = C_f = 6.8\,\mu F$, input current of 1 μA and $V_{DD}$ of 0.6 V.

At rest, the extracellular side of a nerve cell has excess positive charge, and the intracellular side has excess negative charge maintained by the insulating property of the lipid cell membrane which acts as a barrier to charges except at the non-gated and voltage-gated ion channels (Na, K, Ca, and Cl). Thus, the cell's resting potential is negative and is closest to the $K^+$ Nernst potential (−75 mV) as these are the most permeable species at rest. Any influx of $Na^+$ is counterbalanced by the efflux of $K^+$ to maintain the membrane potential constant. As the membrane is depolarized near the action potential threshold, voltage-gated $Na^+$ channels open rapidly, resulting in an increased influx of $Na^+$, which serves as positive feedback, further depolarizing the membrane and moving the potential close to the $Na^+$ Nernst potential (+55 mV), resulting in a spike (Fig. 1e).

Analogous to the operation of nerve cell, a spike is generated in the OECN circuit by integration of the current injected into the input terminal ($I_{in}$). As shown in Fig. 1b, c and f, the integration of the current is carried out by the capacitor $C_{mem}$ which increases the voltage $V_{mem}$ gradually, and upon reaching a specific threshold value, a nerve pulse is fired at $V_{out}$. This is enabled by the noninverting amplifier block A and the positive feedback capacitor $C_f$. The amplifier gain increases rapidly after reaching its transition voltage $V_T$, which is related to the threshold voltage of P($g_4$2T-T) and BBL OECTs. The capacitor $C_{mem}$ can be charged linearly until $V_T$ is reached, but if the current supply is removed before reaching this threshold, the built-up voltage will sink to the ground similar to failed initiations in the biological neuron (Fig. 1e). If $V_{mem}$ reaches $V_T$, the amplifier turns on and the $V_{out}$ increases. A change in the $V_{out}$ value will result in a change in the input voltage $V_{mem}$ enabled by the capacitance-voltage divider circuit implemented by $C_f$ and $C_{mem}$. A small change in the input will further increase the $V_{out}$, leading to a feedback sequence and an exponential increase of voltage, resulting in the action potential generation. A similar positive feedback loop is achieved in a biological neuron by voltage-gated $Na^+$ channels, resulting in further depolarization and opening of even more gated channels causing an exponential increase of cytoplasmic voltage until it is closer to the Nernst potential of Na$^+$. Once the potential reaches this value, Na$^+$ channels close, and K$^+$ channels open, leading to an efflux of K$^+$ and a restoration of the membrane potential to the resting value by the process called repolarization. These processes are repeated in the adjacent regions in the membrane and the action potential is propagated.

The resetting transistor T$_{reset}$ works analogously to the voltage-dependent potassium channels in the nerve cell. When $V_{out}$ is high enough, T$_{reset}$ turns on, and the capacitance $C_{mem}$ discharges through the T$_{reset}$, resulting in a reduction of $V_{mem}$. Once $V_{mem}$ reduces to the threshold voltage of the amplifier $V_T$, the feedback loop is initiated again, resulting in a significant reduction in $V_{out}$ for a slight change in $V_{mem}$. Thus, $V_{out}$ reduces to zero, and the T$_{reset}$ turns off, resulting in termination of the nerve pulse, and the cycle is repeated.

The membrane capacitance plays a crucial role in the speed of the conduction of the action potential in a biological neuron[36]. A lower membrane capacitance results in faster propagation because it is easier to change the potential of the adjacent region as fewer charges are sufficient to change it ($\Delta V = \Delta Q/C$) compared to a membrane with higher capacitance. In nerve cells, this reduction in capacitance is facilitated by wrapping an insulating layer called myelin over the axon. Analogous to this, the OECN spike frequency and width can be modulated by altering the $C_{mem}$ and $C_f$ capacitances in the circuit. Figure 2a, b and Supplementary Figure 3 shows the modulation of the frequency and the full width at half maximum (FWHM) of the spikes of the OECN with capacitance for a constant input current of 1 µA. A lower capacitance value reduces the charging time to reach the spiking threshold voltage and results in higher frequency spikes. With $C_{mem} = C_f = 100$ nF, the firing frequency approaches 0.1 Hz, which matches the low firing rate of certain single neurons[37]. Similarly, the peak width will depend on the discharging time of the capacitors through the resetting transistor T$_{reset}$. Hence, the FWHM will be higher for higher capacitance values. It is noted that there is no further increase in frequencies on reducing the capacitances ($C_{mem} = C_f$) below 100 nF, probably due to limits imposed by the intrinsic overlap capacitances of the printed OECTs.

To further investigate the firing frequency dependence on the OECT performance, we built SPICE models for both p-type and n-type OECTs (Supplementary Figure 5). The simulated transient behaviors of both p-/n-type OECTs were matched with measured results (rise time ($t_r$) ~230–390 ms[34]) by choosing the appropriate value of capacitance C1, as shown in Supplementary Figure 6 and 7a. The simulated neuron spiking frequency reaches 48 mHz, 75 mHz, and 85 mHz for $C_{mem} = C_f = 6.8$ µF, 1 µF, and 100 nF, respectively at an input current of 1 µA (Supplementary Figure 7b–d), showing excellent match with the experimental results (Fig. 2a). Further improving the OECT switching speed, by reducing the channel dimensions, could enable higher firing frequency. For $t_r = 0.5$–1 ms for both p-/n-type OECTs, typical of short-channel photolithographically-made OECTs[38], the simulated firing frequency reaches 95 Hz for $C_{mem} = C_f = 2$ nF and 1 µA current input (Supplementary Figure 8). Such a relatively high and biologically plausible frequency could mimic most neural firing rates (Supplementary Figure 4)[39]. Strategies for reaching such high spiking frequency by reducing the device dimensions are discussed in Supplementary Note 1.

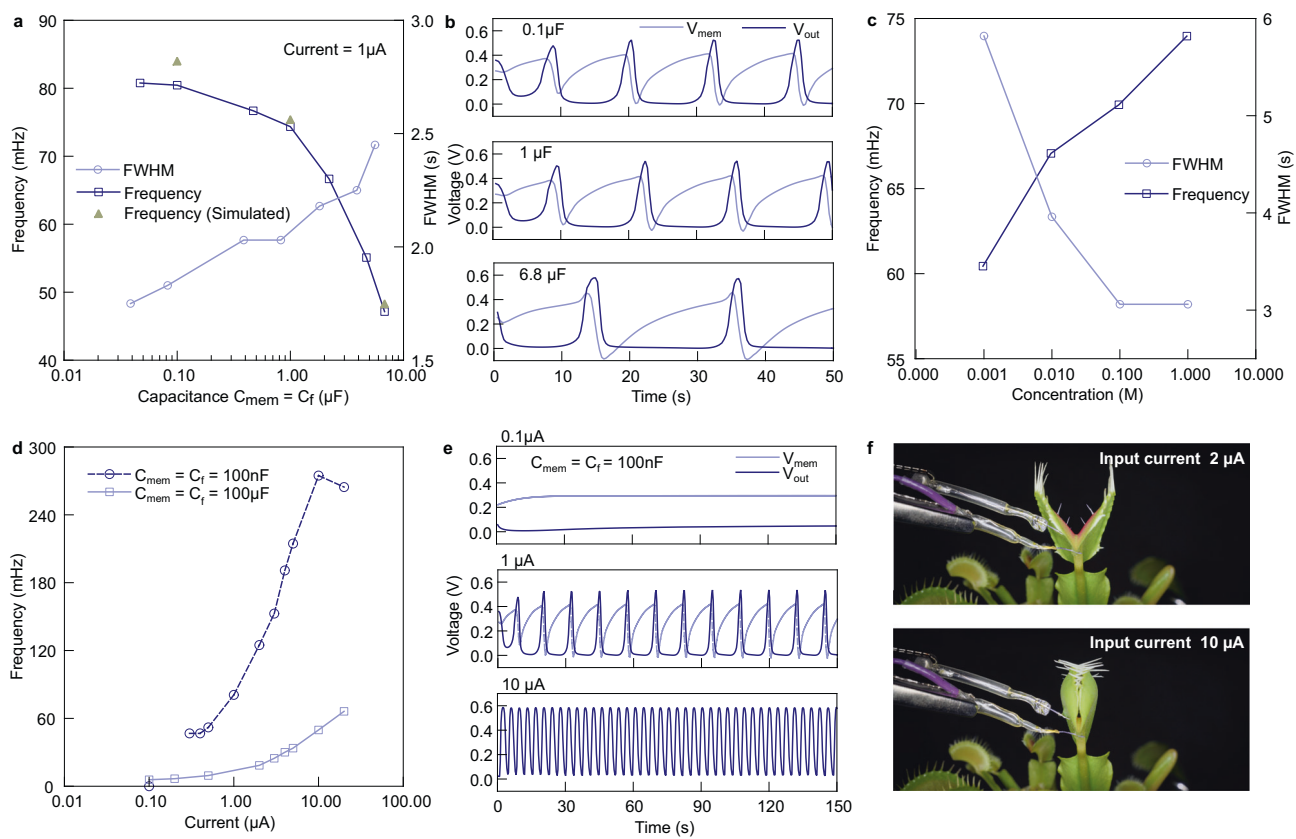

**Fig. 2 Electrical characterization and bio-integration of the organic electrochemical neurons. a** Modulation of frequency and FWHM of spikes with changing $C_{mem}$ and $C_f$ at a constant input current of 1 µA. The 5% deviation in frequency between experimental and simulated values at low capacitances is due to the intrinsic capacitances dominating $C_{mem}$ and $C_f$. **b** Changes of the spiking patterns at 3 different capacitances (0.1, 1, and 6.8 µF). **c** Modulation of spiking frequency and FWHM with various NaCl concentrations. **d** Frequency modulation of neuron with the input current for two different capacitance configurations. **e** Spiking patterns at three different input currents (0.1, 1, and 10 µA) with $C_{mem} = C_f = 100$ nF. **f** Modulation of Venus flytrap using the artificial neuron: the flytrap does not close at a low (2 µA) input current to the neuron but closes when the input current is 10 µA.

A striking feature of this OECN, especially when compared to Si-based or OFET-based spiking neurons, is the ability to control the spiking frequency directly by modulating the ion concentration of the electrolyte. It is observed that the amplifier transition voltage $V_T$ shifts from 0.5 V to 0.2 V as a function of the ion concentration in the electrolyte ($10^{-4}$ to 1 M, Supplementary Figure 9). This shift originates from the threshold voltage dependence of the individual transistors on the ion concentration—the transfer curves of the n-OECTs shift toward lower positive values on increasing the concentration (Supplementary Figure 10), resulting in a shift in $V_T$. A low transition voltage of the amplifier means that the $V_{mem}$ has to reach a lower value to initiate the exponential rise in $V_{out}$ and the positive feedback in the neuron, thus resulting in a higher spiking frequency. The spiking frequency can be increased by 25% on increasing the electrolyte concentration (Fig. 2c). This is also associated with a concurrent reduction in the FWHM of the spikes.

A spiking neuron is generally characterized by its output response for a range of injected currents. The Frequency-Current (F-I) curves of the OECN at two different capacitance configurations are shown in Fig. 2d. Similar to the leaky behavior of the biological neurons that require that the membrane voltage exceeds a given threshold to generate a pulse, the OECN circuit does not fire below a specific current threshold value. In the $C_{mem} = C_f = 100$ nF configuration, a current below 100 nA cannot charge the capacitors enough to reach the spiking threshold (Fig. 2e and Supplementary Figure 11). The frequency can be modulated from 46 mHz at 300 nA to 274 mHz at 10 µA as the capacitors can be charged faster at higher currents bringing the $V_{mem}$ closer to the threshold value faster. Further increase in current does not increase the spiking frequency and is limited by the inverter delay time. The $V_{DD}$ of the inverter can also modulate the spiking frequency of the OECN as shown in Supplementary Figure 12.

Low power consumption is crucial for the application of the circuit in SNNs and event-based sensors. The primary source of power dissipation in this circuit is the amplifying block. Hence, the dynamic power consumption of the circuit is the product of $I_{DD,dynamic}$ and $V_{DD}$. Since the inverter can be operated at a low operating voltage of 0.6 V and the maximum value of $I_{DD}$ dynamic is 25 µA, the maximum dynamic power consumption is 15 µW. This value is lower than the dynamic power consumption of 40 µW reported in OFET-based Axon-Hillock circuit. Furthermore, the power consumption of the OECN can be reduced to much lower values by reducing the channel dimensions by lithographic techniques, which will reduce the current flowing through the OECT. A smaller channel will also increase the response time of the OECT and enable a lower FWHM of the spike to reduce energy consumption.

A monolithically integrated version of the neuron is fabricated by replacing the external capacitors with printed ones based on lateral carbon electrodes (Fig. 1c) and 1 M NaCl as the electrolyte. The characteristics of the fully printed neuron are similar to those discussed above, albeit that it spikes at a slightly lower frequency (Supplementary Figure 13). It is observed that the electrolyte concentration of the capacitor does not change the spiking frequency of the neuron (Supplementary Figure 14). Interestingly, the spiking characteristics remain unchanged even in the absence of the capacitors indicating that the internal capacitance of the circuit induced by the OECTs is larger than the carbon electrode-based capacitors and is sufficient to act as $C_{mem}$, eliminating the need for additional capacitors as required in Si-based circuits.

As a demonstration of the bio-integration capability of the OECN, we interfaced this fully printed neuron with a Venus Flytrap (*Dionaea muscipula*). Venus Flytraps (VFTs) have a thigmonastic response and catch insects by closing the lobes when their mechanosensitive hairs are mechanically elicited. Physical stimulation of the hairs triggers the release of $Ca^{2+}$ in the cytosol[40]. For the trap to close, the cytosolic $Ca^{2+}$ concentration must reach a putative threshold which is typically obtained by stimulating the hair twice within a time interval approximately lower than 30 sec. Trap closure can also be induced by electrical stimulation, including DC stimulation, direct charge injection[41,42], AC stimulation, and capacitive induced current flow[43], making it ideal for integration with the artificial neuron. We used Ag/AgCl electrodes to feed the output of the OECN to the lobes of the VFT with the ground of OECN connected to the lobe and the $V_{out}$ to the midrib. The detailed experimental setup is described in the Methods section and Supplementary Figure 16. Injection of a high current (10 µA) results in a higher frequency (100 mHz) output of the OECN, resulting in the closure of the VFT (Supplementary Video 1). In contrast, a lower current input (2 µA) causes the OECN to spike at a lower frequency (44 mHz) and does not induce closure (Supplementary Video 2, Fig. 2f). We hypothesize the cytosolic $Ca^{2+}$ threshold is met only when the OECN output frequency is high, while at a lower frequency, the cytosolic $Ca^{2+}$ remains below the threshold required to close the VFT. The possibility of modulating the plant's electrophysiology using the OECN opens up new avenues for integration of artificial neuromorphic devices and various biological systems. For example, the unique ability of OECTs to sense multiple biological, physical, and chemical signals enables multiple sensory detection, and their possible fusion in the neuron itself allows the development of novel bio-integrable event-based sensors with sensory fusion. In addition, the ion-based operation mechanism enables facile integration and feedback from biological systems. However, the main advantage of OECNs over other modulation techniques will be the capability for supervised or unsupervised learning at the sensor level. For this, the integration of OECNs with organic artificial synapses is inevitable. Here, we utilize printed organic electrochemical synapses (OECSs) based on electropolymerization mechanism to enable the same, as discussed in the next sections.

**Printed organic electrochemical synapses**. The OECSs are fabricated on the same printed electrode architecture as described for OECNs. The synaptic channel is formed by electropolymerizing a zwitterionic monomer precursor of a conducting polymer (2-(2,5-bis(2,3-dihydrothieno[3,4-b][1,4]dioxin-5-yl)thiophen-3-yl) ethyl(2-(trimethylammonio)ethyl) phosphate, ETE-PC), similar to our previous reports on evolvable OECTs using the anionic sulfonated form of this monomer, ETE-S[12,13] (Fig. 3a, b). The zwitterionic nature of the ETE-PC side chain enables better interaction with the PET substrate compared to ETE-S and results in the formation of a conductive channel when $-0.6$ V is applied to the gate. The OECT characteristics are reported in Supplementary Figure 17. Figure 3a shows the analogy between a biological synapse and the printed OECS. The voltage applied to the Ag/AgCl gate of the OECT represents the presynaptic input and the current response to a presynaptic input, measured between source and drain, is the postsynaptic output referred to as the Excitatory and the Inhibitory Post Synaptic Current (EPSC and IPSC). Like a biological synapse, the OECS operates in 2 modes—short-term and long-term plasticity modes. The ionic accumulation in the poly-ETE-PC channel during the gate-induced doping/dedoping in the absence of ETE-PC monomer results in a short-term increase/decrease in channel conductance on the timescale of several seconds. This process, caused by the accumulation of ions in the channel is similar to the short-term synaptic facilitation (ms to s), which occurs in biological synapses encountering repeated action potentials due to the accumulation of excess $Ca^{2+}$ and neurotransmitters. Figure 3c depicts the evolution of the EPSC with time on the application of gate pulses with a duration of 1 s. The EPSC can be modulated by tuning gate bias from −0.05 V to −0.2 V, showing an excitatory effect for over 100 s. The OECS also exhibits Paired Pulse Facilitation (PPF)/Paired Pulse Depression (PPD) with the peak of the EPSC/IPSC increasing/decreasing on successive application of presynaptic spikes (Fig. 3d). PPF is essential for decoding temporal information in biological systems. PPF indices of the OECS decay exponentially with the pulse intervals and is similar to the signaling characteristics of biological systems (Fig. 3e).

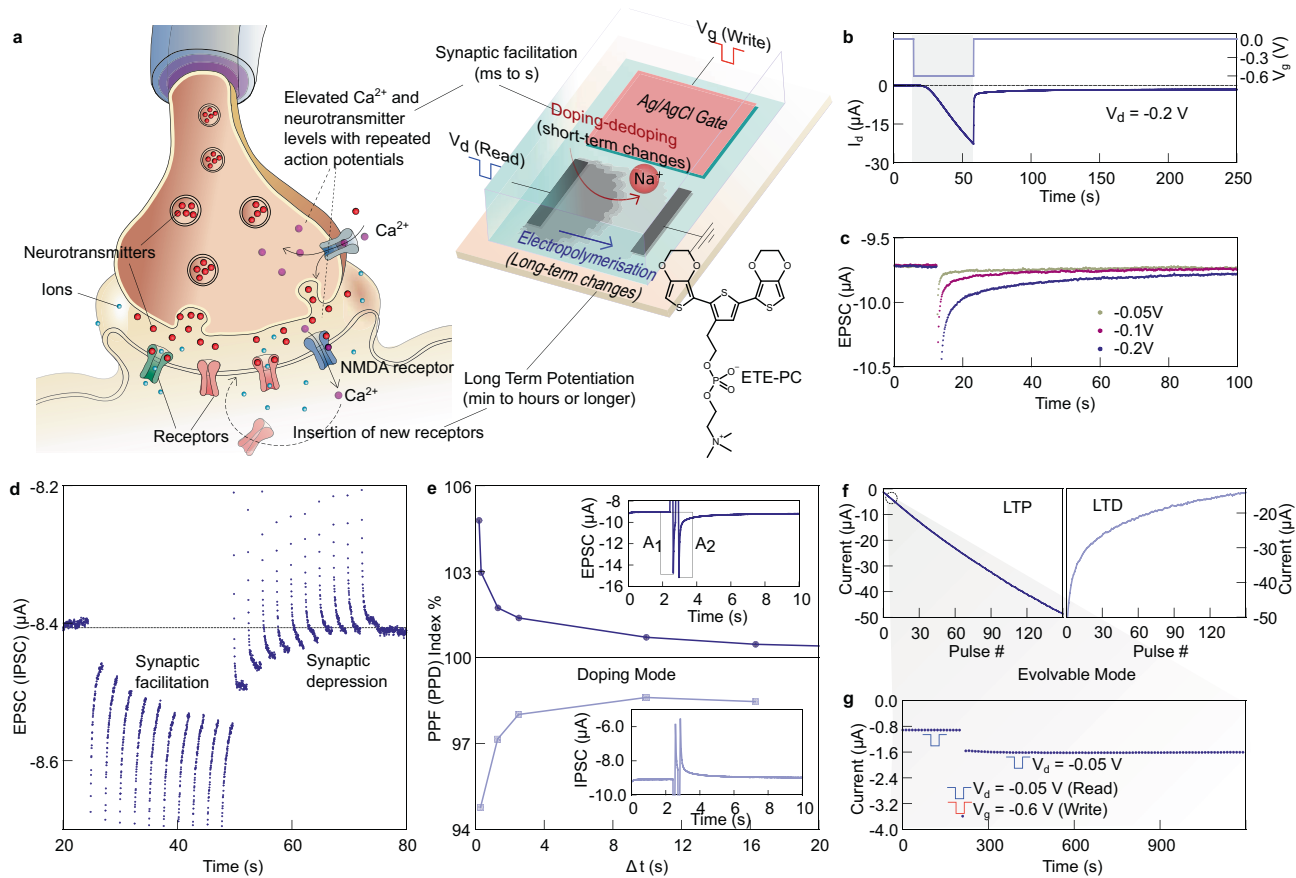

**Fig. 3 Printed organic electrochemical synapses. a** Schematic of the biological and printed organic electrochemical synapses showing analogy between electropolymerization and insertion of new receptors to cause long-term potentiation; and doping-dedoping process and elevation of neurotransmitter levels resulting in synaptic facilitation. The structure of the monomer ETE-PC is also shown. **b** Electropolymerization process to form the semiconducting channel on application of 0.6 V pulse at the gate (drain is kept at −0.2 V and source at 0 V) and resulting increase in drain current after the pulse. **c** Excitatory post synaptic current measured at the drain on application of various gate pulse voltages. **d** Synaptic facilitation and depression with the application of −0.1 V and +0.02 V pulses at the gate showing accumulation of charges in the charges on successive pulses. **e** Paired pulse facilitation and paired pulse depression indices of the synapse operating in the short-term doping-dedoping mode. **f** Long-term potentiation and depression (LTP and LTD) achieved by stepwise electropolymerization (−0.6 V, 1 s pulses) and overoxidation (−2 V, 1 s pulses) of the channel. **g** Long term stability of a single state on electropolymerization showing retention times > 1000 s. The write pulse is −0.6 V (1 s) on the gate and read pulses are −0.05 V (1 s) on the drain applied every 20 s to measure state retention.

A long-term increase in conductivity of OECS is achieved by electropolymerizing additional ETE-PC in the channel through the application of gate voltage pulses of −0.6 V for a duration of 1 s (Fig. 3f). This process is analogous to the N-methyl-D-aspartate (NMDA) receptor-mediated insertion of new receptors in the biological synapse, leading to a long-lasting increase in synaptic strength (minutes to hours and more). A total of 150 distinct states are demonstrated in the OECS, with state retention of more than 1000 s (Fig. 3g). This retention is significantly higher than other OECT-based synapses and eliminates the need for external switches to achieve state retention[15]. Long-term depression (LTD) can be initiated in the synapse by over-oxidizing the channel, which is obtained by applying sufficiently high gate voltage pulses of −2 V for durations of 1 s in the absence of ETE-PC. However, LTD is highly nonlinear compared to a more linear LTP in this OECS. The need to remove the monomer solution to over-oxidize the channel hinders fast reversibility of the process and alternative approaches to induce long-term depression in the system are under investigation (see Supplementary Note 4).

**Integration of organic electrochemical neurons and synapses.** In biological synapses, the synaptic strength does not change for every presynaptic input, as that would quickly saturate the synaptic strength, making any learning impossible. Time correlations between pre- and postsynaptic neuronal spikes form the basis for long-term increases/decreases in synaptic plasticity, known as spike-timing-dependent plasticity (STDP), enabling associative learning. Here, we demonstrate symmetric Hebbian-type STDP with maximum change in synaptic strength when

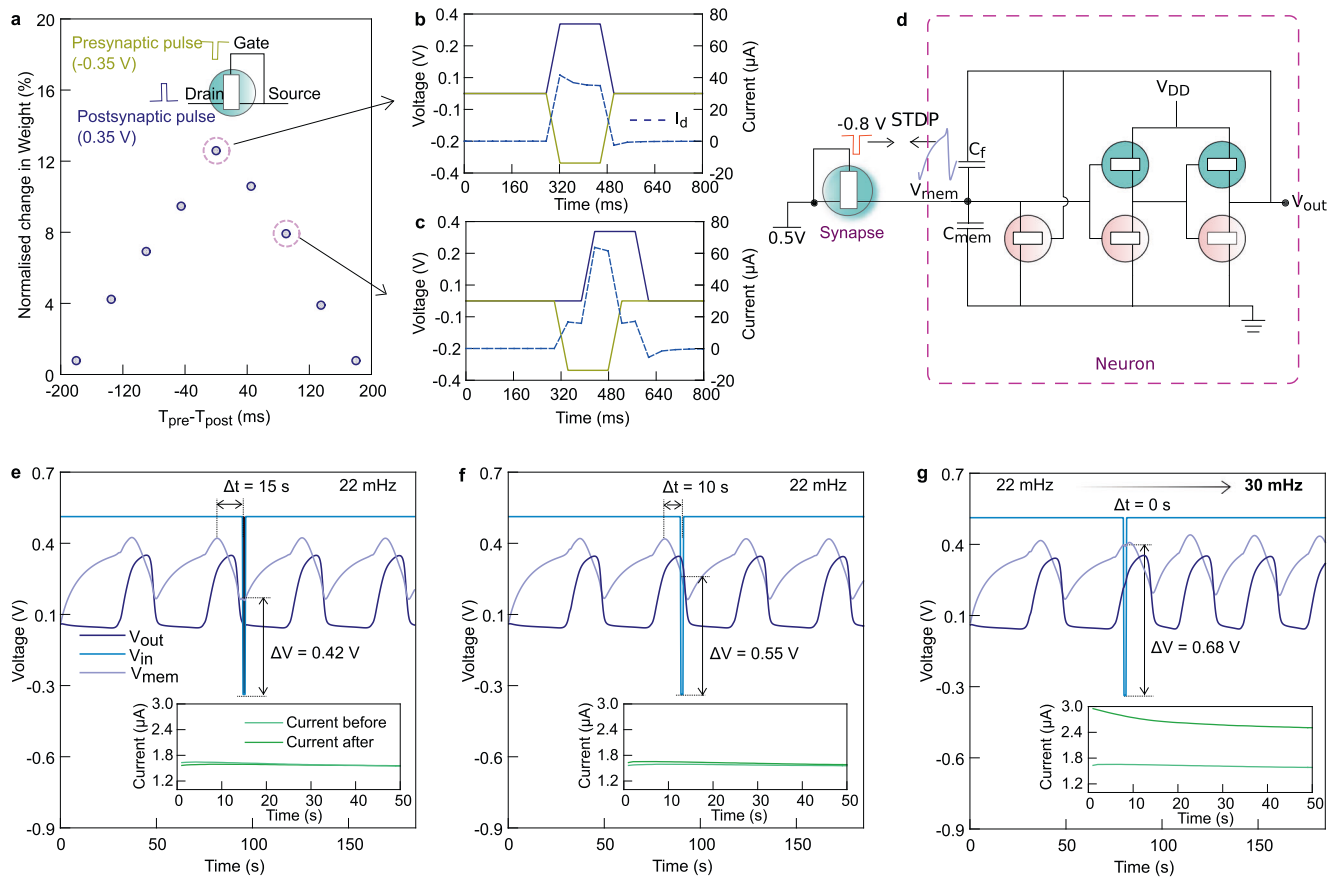

**Fig. 4 Integrated organic electrochemical neurons and synapses. a** Symmetric Hebbian STDP in the artificial synapse. Source and gate electrodes are connected together and form the presynaptic input, while the drain is the postsynaptic output. **b, c** Characteristic voltage waveforms at $\Delta t$ ($T_{pre}$-$T_{post}$) = 0 and 90 ms respectively. **d** Schematic showing the connection of ETE-PC based synapse with the neuron to enable STDP in the system—the presynaptic input is a pulse of $-0.8$ V (1 s) and the $V_{mem}$ is the postsynaptic feedback. **e**–**g** represent the change in synaptic conductivity and the resulting change in frequency of the neuron for $\Delta t$ = 15, 10, and 0 s. $C_{mem}$ and $C_f$ of 100 μF are used in this setup to make the neuron fire at a lower rate for the ease of adjusting the spike delays between pre- and postsynaptic spikes.

pre- and postsynaptic signals coincide and a lower change in strength (conductivity) when the timing of the pulses is mismatched. Figure 4a–c shows the configuration of the pre- and postsynaptic terminals and the characteristic spikes applied to each terminal. Voltage pulses are chosen such that the maximum voltage difference (0.6 V), which is required for electropolymerization, is applied only when both pre- and postsynaptic spikes overlap. The extent of the overlap, which reaches a maximum when the spikes are applied simultaneously, determines the change in channel conductance. The changes in conductance are normalized by the initial channel conductance to present the data in percentage changes in weights. A difference in the timing of the spikes results in a reduced time duration of the spike overlap, and hence lower increase in synaptic strength. This process is analogous to the NMDA receptor-activated insertion of new receptors in synapse leading to LTP; NMDA acts as a biological 'AND' gate and only opens when presynaptic spikes cause glutamate to bind to it, and postsynaptic spike removes the $Mg^{2+}$ blockade. Since electropolymerization can occur even at millisecond time scales, the STDP behavior can be tuned to match biological time scales by optimizing the pulse duration and amplitude.

To further illustrate the significance of the OECN and OECS, we demonstrate a simple neuro-synaptic system with Hebbian learning using a single synaptic transistor connected to the OECN (Fig. 4d). Instead of excitatory current input to the neuron, a voltage is applied to the synapse, which is converted to current based on its synaptic strength, resulting in modulation of the spiking frequency (Supplementary Figure 20). Since the gate and drain of the synaptic ETE-PC-based OECT are kept at a common voltage of 0.5 V, it operates in the fully depleted mode with low conductivity. The presynaptic pulses of magnitude $-0.8$ V and duration of 1 s are applied to the synapse at predefined delays of 0, 5, and 10 s with respect to the time when the value of $V_{mem}$ is maximum. Any delay between the two results in a lower rate of electropolymerization and lower or no change in the spiking of the postsynaptic neuron (Fig. 4e, f). The $V_{mem}$ of the neuron generates the postsynaptic feedback to the synapse. If the presynaptic input arrives in sync with the maximum value of $V_{mem}$, there is a maximum voltage difference across the synapse (0.68 V) for electropolymerization to take place, and the synaptic strength increases significantly. This results in an increase in the firing frequency of the neuron from 22 mHz to 30 mHz, as shown in Fig. 4g, demonstrating the concept of 'neurons which fire together wire together' put forward by Hebb[44]. Si-based systems require complicated circuits with multiple transistors to emulate STDP behavior. Hence, the demonstration of Hebbian learning in this organic electrochemical neuro-synaptic system is an important step and can be extended to build more complicated sensory and processing systems with local learning capabilities.

## Discussion
We have demonstrated the first OECT-based spiking neuron, here made from an all-printed c-OECT technology, and its integration with printed OECT-based artificial synapses exhibiting a range of learning behaviors, including long-term and short-term potentiation and depression and STDP. A neuro-synaptic system with STDP is demonstrated with much fewer elements in comparison to Si-based circuits. The OECT-based circuit can be printed on large scale[31] and with high manufacturing yield[45], which greatly simplifies the production protocol, as the driving strengths of p- and n-type transistors can be matched easily by tuning the thickness of the semiconductor layers. The neuron can be fully printed on flexible substrates and operates at a much lower power compared to OFET-based circuits and can thus enable the development of distributed low-cost smart labels for the future Internet of Things (IoT). The spiking frequency of the OECN can be modulated by changing the input current, membrane capacitance, and voltage input to the amplifier. The properties mentioned above and the ability to tune the frequency of spiking via modulation of electrolyte concentration, which is unique to this OECN, offer facile integration with biological systems that work by similar mechanisms and facilitate the development of future implantable devices. We demonstrate this possibility by interfacing the OECN with a Venus Flytrap to induce the closure of its lobes based on the neuron's firing frequency. Furthermore, we demonstrate the integration of the OECN with an OECS with Hebbian learning capabilities. This, along with the unique ability of the OECTs to sense multiple biological, physical, and chemical signals, enables multiple sensory detection. The possibility to fuse multiple sensing elements in the neuron itself could enable the development of novel bio-integrable event-based sensors for applications ranging from smart neuromorphic labels for IoT packaging and continuous body health monitoring (i.e., wearable electronics) to brain-machine interfaces. Our findings open for the possibility to integrate localized artificial neurosynaptic systems, composed of OECNs and OECSs, with the signaling systems of plants, and with the diffusive, peripheral, and central nervous systems of invertebrates and vertebrates, respectively.

## Methods

**Materials**. The polyethylene terephtalate (PET) substrate Polifoil Bias is purchased from Policrom Screen. Ag 5000, DuPont Silver Ink is used for printed interconnects. Carbon ink 7102 printing paste from DuPont is used for the electrode contacts. Insulating ink (5018, DuPont) is used for electrode isolation. PQ-10, PSSNa, and BBL were purchased from Sigma-Aldrich and used as received. The P(g42T-T) and BBL inks are fabricated through a solvent exchange method. P($g_42$T-T) (30 mg) dissolved in chloroform (15 mL) is added dropwise to iso-propanol (75 mL) under high speed stirring (1500 rpm) to obtain bright blue nanoparticles of P($g_42$T-T). BBL (30 mg) dissolved in methanesulfonic acid (15 mL) is added dropwise to isopropanol (75 mL) under high-speed stirring (1500 rpm) to obtain dark purple BBL nanoparticles. The P($g_42$T-T) and BBL nanoparticles are respectively collected by centrifugation (5000 rpm, 30 min) and washed in IPA six times until neutral. The neutral P($g_42$T-T) and BBL nanoparticles are re-dispersed in IPA to obtain dispersion inks (about 0.006 mg/mL for P($g_42$T-T) and 0.1 mg/mL for BBL)[34,46].

**Fabrication of OECTs**. Flatbed sheet-fed screen-printing equipment (DEK Horizon 03iX) is used to deposit all materials on top of PET plastic substrates. To form the contacts, a layer of carbon is first deposited on the substrate. This is followed by the Ag 5000 silver ink deposition and the insulating layer of 5018 ink, which is UV-cured to define the channel and the gate. Ag/AgCl ink is blade coated through a shadow mask and annealed at 120 ºC for 10 min to form the gate. The P(g42T-T) and BBL layers are deposited by spray-casting in air through a shadow mask, using a standard HD-130 air-brush (0.3 mm) at an atomization air pressure of 1 bar[34].

**Electrical characterization**. All the OECT characterizations are carried out in a semiconductor parameter analyzer (Keithley 4200 SCS) at a temperature of around 20 °C and a relative humidity of around 45%.

**Simulation of OECTs and OECNs**. The SPICE models of p-type and n-type OECTs are developed in B2 SPICE (EMAG Technologies Inc.). Both models are built based on the measured transfer characteristics and transient switching characteristics of printed OECTs. The simulation of OECNs gives the membrane voltage ($V_{mem}$) and the output of the Amplifier A ($V_{out}$) in the OECN circuits.

**Venus Flytrap measurement setup**. Venus Flytraps (*Dionaea muscipula*, VFTs) were purchased from Plantagen (Norrköping, Sweden) and kept in greenhouse conditions (day/night temperature of 28/22.5 °C, 12 h photoperiod, 60% relative humidity, 400 ppm $CO_2$). Ag/AgCl electrodes were prepared by coating two Ag wires (AGW1030, World Precision Instruments) with AgCl and insulating them with Teflon heat shrink tubing, leaving the electrode tips open. To electrically stimulate the VFT, the positive stimulation electrode was placed on the midrib and the ground or negative electrode was attached on the lobe, via a conductive gel (SignaGel, Parker Laboratories, Inc., NJ, USA) to ensure stable electrical contact. The positive stimulation electrode was electrically integrated with the output of the OECN and the two ground terminals were connected together to close the circuit. The Action Potential signals at different frequencies were then supplied by the OECN to induce the VFT trap closure.

## References
1. Mead, C. Analog VLSI and neural systems. 416 (1989).
2. Grahn, P. J. et al. Restoration of motor function following spinal cord injury via optimal control of intraspinal microstimulation: toward a next generation closed-loop neural prosthesis. *Front. Neurosci.* **8**, 296 (2014).
3. Bonifazi, P. et al. In vitro large-scale experimental and theoretical studies for the realization of bi-directional brain-prostheses. *Front. Neural Circuits* **7**, 40 (2013).
4. Azghadi, M. R. et al. Hardware Implementation of Deep Network Accelerators Towards Healthcare and Biomedical Applications. *IEEE Trans. Biomed. Circuits Syst.* **14**, 1138–1159 (2020).
5. Keene, S. T. et al. A biohybrid synapse with neurotransmitter-mediated plasticity. *Nat. Mater.* **19**, 969–973 (2020).
6. Donati, E., Krause, R. & Indiveri, G. Neuromorphic Pattern Generation Circuits for Bioelectronic Medicine. in *2021 10th International IEEE/EMBS Conference on Neural Engineering (NER)* 1117–1120 https://doi.org/10.1109/NER49283.2021.9441285 (2021).
7. Corradi, F. & Indiveri, G. A Neuromorphic Event-Based Neural Recording System for Smart Brain-Machine-Interfaces. *IEEE Trans. Biomed. Circuits Syst.* **9**, 699–709 (2015).
8. John, R. A. et al. Self healable neuromorphic memtransistor elements for decentralized sensory signal processing in robotics. *Nat. Commun.* **11**, 4030 (2020).
9. Indiveri, G. et al. Neuromorphic silicon neuron circuits. *Front. Neurosci.* **5**, 73 (2011).
10. Abu-Hassan, K. et al. Optimal solid state neurons. *Nat. Commun.* **10**, 5309 (2019).
11. Cruz-Albrecht, J. M., Yung, M. W. & Srinivasa, N. Energy-Efficient Neuron, Synapse and STDP Integrated Circuits. *IEEE Trans. Biomed. Circuits Syst.* **6**, 246–256 (2012).
12. Gerasimov, J. Y. et al. An Evolvable Organic Electrochemical Transistor for Neuromorphic Applications. *Adv. Sci.* **6**, 1801339 (2019).
13. Gerasimov, J. Y. et al. A Biomimetic Evolvable Organic Electrochemical Transistor. *Adv. Electron. Mater.* **7**, 2001126 (2021).
14. van de Burgt, Y. et al. A non-volatile organic electrochemical device as a low-voltage artificial synapse for neuromorphic computing. *Nat. Mater.* **16**, 414–418 (2017).
15. Fuller, E. J. et al. Parallel programming of an ionic floating-gate memory array for scalable neuromorphic computing. *Science* **364**, 570–574 (2019).
16. Tuchman, Y. et al. Organic neuromorphic devices: Past, present, and future challenges. *MRS Bull.* **45**, 619–630 (2020).
17. Ji, X. et al. Mimicking associative learning using an ion-trapping non-volatile synaptic organic electrochemical transistor. *Nat. Commun.* **12**, 2480 (2021).
18. Kim, Y. et al. A bioinspired flexible organic artificial afferent nerve. *Science* **360**, 998–1003 (2018).

19. Seo, D.-G., Go, G.-T., Park, H.-L. & Lee, T.-W. Organic synaptic transistors for flexible and stretchable artificial sensory nerves. *MRS Bull.* **46**, 321–329 (2021).
20. Khodagholy, D. et al. In vivo recordings of brain activity using organic transistors. *Nat. Commun.* **4**, 1575 (2013).
21. Cea, C. et al. Enhancement-mode ion-based transistor as a comprehensive interface and real-time processing unit for in vivo electrophysiology. *Nat. Mater.* **19**, 679–686 (2020).
22. Benfenati, V. et al. A transparent organic transistor structure for bidirectional stimulation and recording of primary neurons. *Nat. Mater.* **12**, 672–680 (2013).
23. Lanzani, G. Organic electronics meets biology. *Nat. Mater.* **13**, 775–776 (2014).
24. Higgins, S. G., Fiego, A. L., Patrick, I., Creamer, A. & Stevens, M. M. Organic Bioelectronics: Using Highly Conjugated Polymers to Interface with Biomolecules, Cells, and Tissues in the Human Body. *Adv. Mater. Technol.* **5**, 2000384 (2020).
25. Berggren, M. et al. Ion Electron–Coupled Functionality in Materials and Devices Based on Conjugated Polymers. *Adv. Mater.* **31**, 1805813 (2019).
26. Paulsen, B. D., Fabiano, S. & Rivnay, J. Mixed Ionic-Electronic Transport in Polymers. *Annu. Rev. Mater. Res.* **51**, 73–99 (2021).
27. Hosseini, M. J. M. et al. Organic electronics Axon-Hillock neuromorphic circuit: towards biologically compatible, and physically flexible, integrate-and-fire spiking neural networks. *J. Phys. Appl. Phys.* **54**, 104004 (2020).
28. Romele, P. et al. Multiscale real time and high sensitivity ion detection with complementary organic electrochemical transistors amplifier. *Nat. Commun.* **11**, 3743 (2020).
29. Rivnay, J. et al. Organic electrochemical transistors. *Nat. Rev. Mater.* **3**, 1–14 (2018).
30. van de Burgt, Y., Melianas, A., Keene, S. T., Malliaras, G. & Salleo, A. Organic electronics for neuromorphic computing. *Nat. Electron.* **1**, 386–397 (2018).
31. Andersson Ersman, P. et al. All-printed large-scale integrated circuits based on organic electrochemical transistors. *Nat. Commun.* **10**, 5053 (2019).
32. Andersson Ersman, P. et al. Screen printed digital circuits based on vertical organic electrochemical transistors. *Flex. Print. Electron.* **2**, 045008 (2017).
33. Hütter, P. C., Rothländer, T., Scheipl, G. & Stadlober, B. All Screen-Printed Logic Gates Based on Organic Electrochemical Transistors. *IEEE Trans. Electron Devices* **62**, 4231–4236 (2015).
34. Yang, C.-Y. et al. Low-power/high-gain flexible complementary circuits based on printed organic electrochemical transistors. *Adv. Electron. Mater.* **n/a**, 2100907 (2022).
35. Purves, D. *Neuroscience.* (Sinauer Associates, Sunderland, 2004).
36. Kandel, E. R., Schwartz, J. H., Jessell, T. M., Siegelbaum, S. A. & Hudspeth, A. J. *Principles of Neural Science.* (McGraw-Hill, New York, 2013).
37. Kerr, J. N. D., Greenberg, D. & Helmchen, F. Imaging input and output of neocortical networks in vivo. *Proc. Natl Acad. Sci.* **102**, 14063–14068 (2005).
38. Wu, H.-Y. et al. Influence of Molecular Weight on the Organic Electrochemical Transistor Performance of Ladder-Type Conjugated Polymers. *Adv. Mater.* **34**, 2106235 (2022).
39. Baddeley, R. et al. Responses of neurons in primary and inferior temporal visual cortices to natural scenes. *Proc. R. Soc. Lond. B Biol. Sci.* **264**, 1775–1783 (1997).
40. Suda, H. et al. Calcium dynamics during trap closure visualized in transgenic Venus flytrap. *Nat. Plants* **6**, 1219–1224 (2020).
41. Volkov, A. G., Adesina, T. & Jovanov, E. Charge induced closing of Dionaea muscipula Ellis trap. *Bioelectrochemistry* **74**, 16–21 (2008).
42. Volkov, A. G., Adesina, T. & Jovanov, E. Closing of venus flytrap by electrical stimulation of motor cells. *Plant Signal. Behav.* **2**, 139–145 (2007).
43. Li, W. et al. An on-demand plant-based actuator created using conformable electrodes. *Nat. Electron.* **4**, 134–142 (2021).
44. Hebb, D. O. *The organization of behavior; a neuropsychological theory*. xix, 335 (Wiley, 1949).
45. Zabihipour, M. et al. High yield manufacturing of fully screen-printed organic electrochemical transistors. *Npj Flex. Electron.* **4**, 1–8 (2020).
46. Kroon, R. et al. Polar Side Chains Enhance Processability, Electrical Conductivity, and Thermal Stability of a Molecularly p-Doped Polythiophene. *Adv. Mater.* **29**, 1700930 (2017).

## Acknowledgements
This work was financially supported by the Knut and Alice Wallenberg foundation, the Swedish Research Council (2016-03979, 2018-06197, and 2020-03243), the Swedish Foundation for Strategic Research (RMX18-0083, FFL18-0101), ÅForsk (18-313 and 19-310), Olle Engkvists Stiftelse (204-0256), VINNOVA (2020-05223), the European Research Council (834677 "e-NeuroPharma" ERC-2018-ADG), the European Commission through the FET-OPEN project MITICS (GA-964677), and the Swedish Government Strategic Research Area in Materials Science on Functional Materials at Linköping University (Faculty Grant SFO-Mat-LiU 2009-00971). R.O. and D.B. are supported by MultiPark, a strategic research area at Lund University.

## Author contributions
P.C.H. and S.F. conceived and designed the experiments. P.C.H. and C.-Y.Y. performed the fabrication and characterization of the OECNs and analyzed the data. C.-Y.Y. developed the P(g₄2T-T)/BBL inks for printing. D.T. designed the circuit, conducted the simulations, and designed the layout for printing. J.Y.G. and P.C.H. fabricated the OECSs and integrated it with OECNs. A.M.D., A.A.-M, P.C.H., and E.S. integrated the OECNs with Venus Flytraps. M.M. printed the circuits. R.K. synthesized P(g₄2T-T). D.B. and R.O. designed and synthesized ETE-PC monomer. P.C.H., D.T., J.Y.G., M.B., and S.F. wrote the manuscript. All authors contributed to discussion and manuscript preparation.

## Funding
Open access funding provided by Linköping University.

## Competing interests
C.-Y.Y., M.B., and S.F. are the founder of n-Ink AB. The other authors declare no competing interests.

## Additional information
**Supplementary information** The online version contains supplementary material available at https://doi.org/10.1038/s41467-022-28483-6.

**Correspondence** and requests for materials should be addressed to Simone Fabiano.

**Peer review information** *Nature Communications* thanks Pierre Leleux, Karl Leo, and Sébastien Sanaur for their contribution to the peer review of this work. Peer reviewer reports are available.

**Reprints and permission information** is available at http://www.nature.com/reprints

**Publisher's note** Springer Nature remains neutral with regard to jurisdictional claims in published maps and institutional affiliations.